\documentstyle[epsfig, twocolumn]{iso98}

\newcommand{\thepapername}{}

\begin{document}

\setlength{\parindent}{0pt}
\setlength{\parskip}{ 10pt plus 1pt minus 1pt}
\setlength{\hoffset}{-1.5truecm}
\setlength{\textwidth}{ 17.1truecm }
\setlength{\columnsep}{1truecm }
\setlength{\columnseprule}{0pt}
\setlength{\headheight}{12pt}
\setlength{\headsep}{20pt}
\pagestyle{veniceheadings}

\title{\bf MODELLING THE INFRARED CONTINUUM OF CENTAURUS A}

\author{{\bf D.M.~Alexander$^1$, A.~Efstathiou$^2$, J.H.~Hough$^3$, 
D.~Aitken$^3$, D.~Lutz$^4$, P.~Roche$^5$, E.~Sturm$^4$} \vspace{2mm} \\
$^1$SISSA, via Beirut 2-4, 34014 Trieste, Italy \\ 
$^2$Astrophysics Group, Imperial College, London, UK \\ 
$^3$Department of Physical Sciences, University of Hertfordshire, Hatfield,
Hertfordshire, UK \\
$^4$Max-Planck-Institut fur extraterrestrische Physik, Garching, Germany \\ 
$^5$Department of Astrophysics, Oxford University, Oxford, UK. }

\maketitle

\begin{abstract}

We present ISOPHOT\_S, ISOSWS and 8 to 13 micron ground based observations
of Centaurus A that show prominent PAH and silicate features. These and
other data are used to construct a model for the infrared continuum. We
find that in a nuclear sized aperture (\verb!~!4 arcsec,\verb!~!60 pc) the
SED is characteristic of emission from a starburst and AGN torus; in
larger apertures an additional component of cirrus emission is required.
Based on our model, the torus diameter is estimated to be 3.6 pc and the
best fitting inclination angle of the torus is 45 degrees. This result has
implications for the detectability of tori in low power AGN and in
particular for the use of the IRAS 60/25 micron flux ratio as an indicator
of the torus inclination. \vspace {5pt} \\


  Key~words: active-galaxies; individual; Cen 
A; nuclei-galaxies; galaxies-radiative transfer.

\end{abstract}

\section{INTRODUCTION}

Rowan-Robinson and Crawford (1989) attribute the infrared (IR) spectra of
galaxies to a mixture of up to 3 components: i) general disc emission from
grains heated by the interstellar radiation field (cirrus), ii) a Seyfert
component peaking in the mid-infrared and iii) a starburst component peaking
at about 60 microns. Essentially all of these sources arise from the thermal
reprocessing of ultra-violet and other high frequency photons by the dust
within these objects. The dust grains re-radiate the absorbed energy in the
IR, with the resultant spectrum dependent on the distribution of dust grain
temperatures. For the starburst and Seyfert components, the clouds of dust
are optically thick, even to IR photons, so radiative transfer effects are
important. 

Centaurus A, the famous southern radio (FRI) counterpart to NGC5128,
identified by Bolton, Stanley and Slee (1949), at a distance of
approximately 3.1 Mpc (Tonry and Schechter, 1990) is the closest active
galaxy to us.  It is a multi-faceted object, showing evidence of a merger
(see Mirabel et al, these proceedings), starburst and AGN activity, with HII
regions, shells, jets, optical filaments and a warped dust lane.

Evidence of star formation is most apparent in the dust lane that
intersects the host galaxy NGC5128. Marston and Dickens (1988) found that
their 12 micron IRAS DSD observations followed the H$\alpha$ emission,
along the dust lane, and hence the regions of star formation. They
modelled their 12, 25, 60 and 100 micron observations as a cirrus spectrum
of small and large grains heated by the interstellar radiation with two
grain temperatures - hot (240K)  small grains and cooler (30K) large
grains. Evidence of starburst activity comes from enhanced far-IR and
sub-mm nuclear emission that shows a structure offset from the dust lane
(e.g.\ Hawarden et al, 1993) and from the high star formation rate, which
is typical of a starburst galaxy (Eckart et al, 1990). Evidence of AGN
activity comes from the radio jets/lobes (e.g.\ Clarke, Burns and Norman,
1992) and variable x-ray emission (e.g.\ Morini, Anselmo and Molteni,
1989) whilst evidence of a Seyfert-like dusty torus comes from the high
optical depth to the nucleus (e.g.\ Blanco, Ward and Wright, 1990). 

\section{OBSERVATIONS AND DATA}

The observations used to constrain the models are an array of published
and unpublished data, taken from a variety of sources and grouped into 3
aperture sizes: a small nuclear aperture of \verb!~!4 arcsec, an
intermediate aperture of \verb!~!20 arcsec and a large aperture of
\verb!~!90 arcsec.

The ISOPHOT\_S observations (intermediate aperture) have not been
presented before, see figure 1. The observations have been reduced with
PIA without additional modifications. The zodiacal light contributions
have been removed using a chopped measurement. The ISOPHOT\_S spectrum
consists of two parts - PHT\_SL (6 to 12 microns) and PHT\_SS (2.5 to 5.5
microns). Clearly seen in the spectrum are the deep 9.7 micron silicate
absorption feature and PAH emission features, however importantly, PAH
emission at 3.3 microns is not clearly detected.

\begin{figure}[!h]
  \begin{center}
    \leavevmode
  \centerline{\epsfig{file=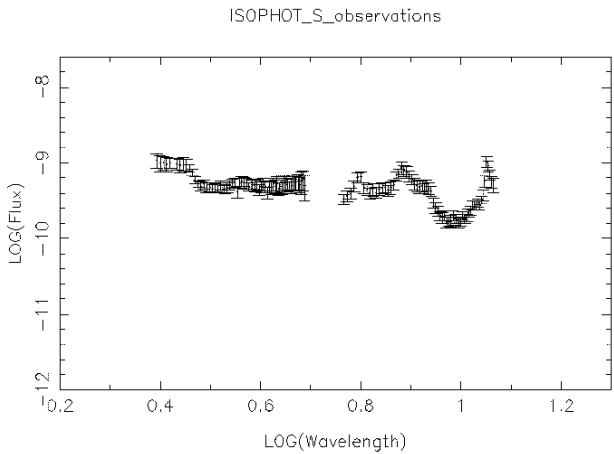,width=8cm}}
  \vspace{0.5cm}
  \end{center}
  \caption{\em ISOPHOT\_S observations. The flux units are
ergs$^-$$^1$cm$^-$$^2$. The wavelength units are LOG(microns).}
  \label{fig:sample1}
\end{figure}

The ISOSWS observations (intermediate aperture) presented here are the
continuum points. The emission line spectrum, details of the observations
and data reduction will be presented in Sturm et al (in preparation). The
continuum points were determined by a second order polynomial fit to the
line free regions of the data. The error bars represent the uncertainty in
flux calibration - the RMS noise level is insignificant by comparison. No
correction has been made for the different aperture sizes at the different
wavelengths. 

Due to the problem of starlight contamination, and some uncertainty in the
flux levels at the short wavelength end of these ISO spectra, the data
shortward of 6 microns are not used in the model fitting.

The UCL spectrometer observations (nuclear aperture) have not been
presented before, see figure 2. The silicate feature at 9.7 microns is
clearly detected. In addition there appears to be a component of 11.3
micron PAH and 12.8 micron $[NeII]$ emission.

\begin{figure}[!h]
  \begin{center}
    \leavevmode
  \centerline{\epsfig{file=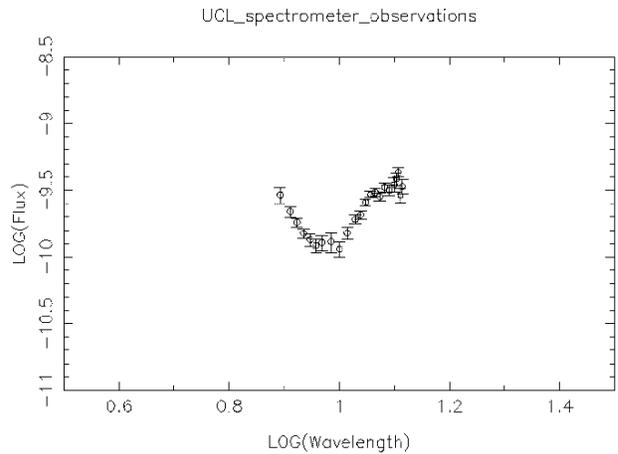,width=8cm}}
  \vspace{0.5cm}
  \end{center}
  \caption{\em UCL spectrometer observations. The units are the same as
in figure 1.}
  \label{fig:sample2}
\end{figure}

Additional data was obtained from the literature. The near-IR points
(nuclear aperture) were taken from Packham et al (1996) and have had the
stellar contribution removed. The sub-mm points (intermediate and large
aperture) were taken from Hawarden et al (1993) and have been corrected
for non-thermal contributions. The 12, 25, 60 and 100 micron large
aperture points were taken from Marston and Dickens (1988)  and are IRAS
DSD observations.

\section{THE MODEL COMPONENTS}

The proposed model for Centaurus A is a combination of a number of
components. The first component is the AGN dusty torus, based on the
NGC1068 tapered disc infrared model of Efstathiou, Hough and Young (1995) 
which assumed a quasar-like central continuum source illuminating an
optically and geometrically thick dusty torus. The main model parameters
are: opening half-angle of the toroidal cone ($\theta$), ratio of the
inner and outer radii (r1/r2), ratio of the height to outer radius (h/r2),
equatorial optical depth ($e_{{\tau}uv}$), dust sublimation temperature
(T1) and radial dependence of the density distribution ($r^{-\beta}$). The
second and third components are the starburst and cirrus models of
Efstathiou, Rowan-Robinson and Siebenmorgan (2000). The cirrus spectrum is
assumed to arise from 12,000 K stars centrally illuminating a diffuse
optically and geometrically thin dust shell. The starburst is modelled as
an ensemble of giant molecular clouds centrally illuminated by hot stars. 
The stellar population is modelled in terms of the population synthesis
models of Bruzual and Charlot (1993) and the giant molecular clouds evolve
with time due to the expansion of the HII regions. Additional information
on the model components is given in Alexander et al, 1999.

To take into account of dust lane absorption, radiation from the torus and
starburst are visually extincted by 10 mags, essentially the same as the
value adopted by Packham et al (1996). The cirrus is not extincted since
it is itself associated with the dust lane. The extinction model used
assumes the grain model and interstellar extinction curve of
Rowan-Robinson (1992).

\section{THE CENTAURUS A MODEL}

For small apertures, and at IR wavelengths below 12 microns, it was
assumed that the SED would be dominated by radiation from the torus, seen
through the dust lane. Therefore the torus model is fitted to this
wavelength range, which includes the 9.7 micron silicate feature. In
fitting to larger apertures, a contribution from the smaller apertures is
included and any additional components required.

The torus parameters were set to be the same as those in the Efstathiou,
Hough and Young (1995) model for NGC1068 ($\theta$=30 degrees, r1/r2=0.01, 
h/r2=0.1, $e_{{\tau}uv}$=1,200, T1=950 and $\beta$=-1) and the SED was
calculated for different values of inclination (the angle between the
polar axis and the line of sight). No value of inclination gave a
simultaneously good fit to both the near-IR continuum and the silicate
feature, with the former best fit by 40 degrees and the latter by 70
degrees. The observed silicate feature shows a feature at 11.3 microns
which is probably PAH emission, therefore a component of starburst was
added. This provided an excellent fit to the silicate feature, with
roughly equal components of torus and starburst for wavelengths less than
60 microns, with the torus at an inclination of 45 degrees, see figure 3. 

For larger apertures a component of cirrus emission was required. For
completeness, the contribution this would make in the nuclear aperture is
shown in figure 3 (the cirrus component was determined assuming constant
surface brightness, as implied from the fits to the larger apertures).

\begin{figure}[!h]
  \begin{center}
    \leavevmode
  \centerline{\epsfig{file=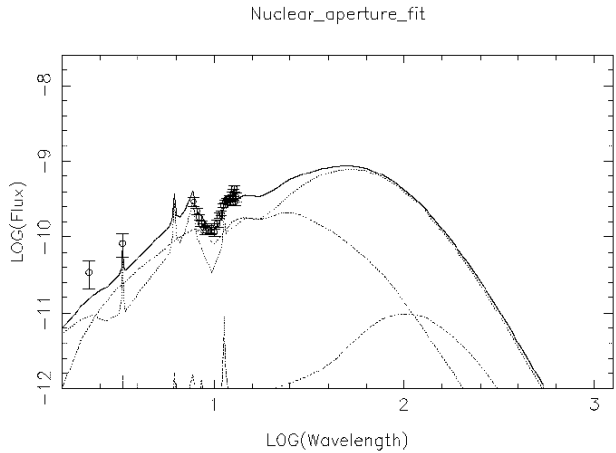,width=8cm}}
  \vspace{0.5cm}
  \end{center}
  \caption{\em The model fit to the nuclear aperture data points. At
the 100 micron wavelength: the lowest intensity model component is the
cirrus, the highest intensity component is the starburst and the
component in between is the torus. The units are the same as in figure 1.}
  \label{fig:sample3}
\end{figure}

In the intermediate aperture, the torus component is kept fixed and an
additional contribution from the starburst is added, producing a
reasonable fit to the intermediate aperture data points for the silicate
feature and IR continuum below 25 microns. The PAH features and the
continuum at longer wavelengths showed that an additional component was
required. Observations (e.g.\ Marston and Dickens, 1988) suggest that this
additional component is cirrus emission from stars within the dust lane.
By slightly reducing the contribution from the starburst and adding a
component of cirrus, an excellent fit to all the data points is obtained,
see figure 4.

\begin{figure}[!h]
  \begin{center}
    \leavevmode
  \centerline{\epsfig{file=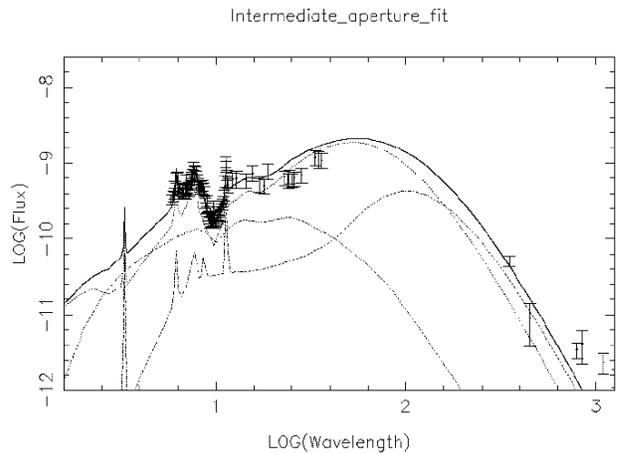,width=8cm}}
  \vspace{0.5cm}
  \end{center}
  \caption{\em The model fit to the intermediate aperture data points. The
model components and units are the same as in figure 3.}
  \label{fig:sample4}
\end{figure}

The observed size of the starburst component (e.g.\ Hawarden et al, 1993)
limits the contribution it should have in the large aperture. Whilst there
should be an additional contribution from the starburst, the cirrus
emission, which is associated with the dust lane, and so is widespread,
should contribute significantly more. The best fit is obtained with a
relatively small increase in the starburst contribution and a significant
increase in the cirrus contribution, see figure 5.

\begin{figure}[!h]
  \begin{center}
    \leavevmode
  \centerline{\epsfig{file=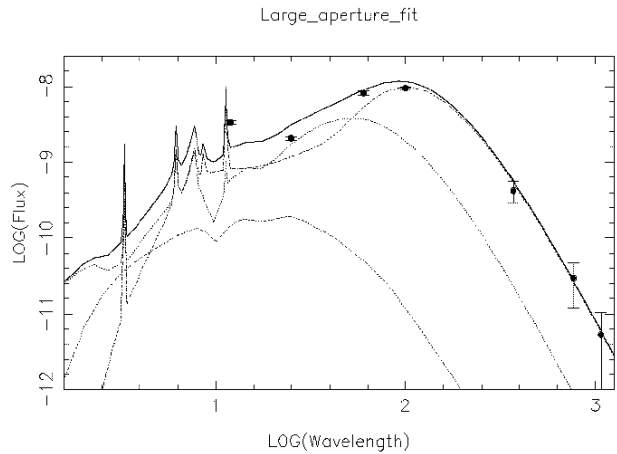,width=8cm}}
  \vspace{0.5cm}
  \end{center}
  \caption{\em The model fit to the large aperture data points. The
model components and units are the same as in figure 3.}
  \label{fig:sample5}
\end{figure}

\section{DISCUSSION}

The complete model includes contributions from torus, starburst and cirrus
in all apertures. The model fit in the nuclear aperture is quantitatively
similar to that found by Laurent et al (these proceedings), with ISOCAM
CVF observations and AGN and starburst observational templates.

The opening half-angle of the torus cone is very similar to that of the
'cones' in the recently published HST nuclear Pa$\alpha$ image (Schreier
et al, 1998), and if it is assumed that these observations are of
ionisation cones, rather that the suggested warped accretion disc, the
cone opening half-angle is constrained between 30 and 40 degrees.
Extinction to the 2.2 micron region can be estimated by comparing the flux
for the face-on torus to the 45 degree inclined torus and is 2.6 mags,
equating to a visual extinction of 23 mags. This degree of extinction is
consistent with the analysis of Meadows and Allen (1992) and suggests that
the near-IR emission region is considerably less extincted than the x-ray
region (70 mags, Blanco, Ward and Wright, 1990). 

Based on this model, the torus diameter is calculated to be 3.6 pc,
assuming a distance of 3.1 Mpc, placing it easily within the nuclear
aperture. This is significantly smaller than that actually observed in
other galaxies (e.g.\ \verb!~!1 kpc for MG0414+0534 (Oya et al, these
proceedings),\verb!~!200pc for NGC1068 (Young et al, 1996), \verb!~!100
pc for NGC4261 (Jaffe et al, 1993) and \verb!~!50 pc for NGC4151 (Mundel
et al, 1995)).  However, although of a small diameter, this model does not
support the even smaller diameter tori of the Pier and Krolik (1992)
model.

Possible observational support for the small torus size estimate comes
from the 2.3 GHz radio observations of the nuclear sub-arcsecond scale jet
and counter jet (Jones et al, 1996). These images show the core to be
completely absorbed between the jet and counter jet, attributed to a
gaseous disc or torus of 0.4 to 0.8 pc. The model torus, with an
inclination of 45 degrees would present a 1.4 pc absorption band; very
similar to the size of the radio core absorption.

Within the context of unified theories the inclination of the torus is a
crucial parameter in determining whether an AGN is a Type 1 or Type 2.
Various studies have suggested that the IRAS 60/25 micron flux ratio is an
indicator of torus inclination with high or low ratios implying edge-on to
face-on torus inclinations. The combination of all the model components
gives a 60/25 micron flux ratio of 9.4, although the actual torus 60/25
micron flux ratio is only 0.7. The torus inclination implied from this
study is identical to that found for NGC1068 (Efstathiou, Hough and Young,
1995) and yet the IRAS 60/25 micron flux ratio for NGC1068 is only 2.1. If
the torus was as powerful as that in NGC1068, or a smaller 'nuclear' sized
aperture was used, the 60/25 micron flux ratio would be more indicative of
the torus inclination but within the large IRAS sized aperture, the torus is
dominated by other more powerful IR emitting components.
 
A comparison of the mean IRAS 60/25 micron flux density ratios of the
extended 12 micron sample of galaxies, which is considered statistically
complete on 12 micron flux, gives a mean of 4.1$\pm$2.9 for the Seyfert 1s
and 4.5$\pm$3.0 for the Seyfert 2s (Andy Thean, private communication). 
This suggests that this colour ratio cannot distinguish between Type 1 and
Type 2 AGN and so cannot be a reliable indicator of the torus inclination.

The IRAS 60/25 micron flux density ratio is more likely to provide a ratio
of the AGN to star formation components. Evidence for this comes from the
60 micron to 6 cm flux correlation (Wilson, 1988)  which works only if the
radio jet/lobe AGN emission is removed (Peter Bartel, private
communication), thereby strongly suggesting that the 60 micron flux is
coming from star formation and starburst regions. Additional evidence
comes from the ISOLWS observations of NGC1068 (Spinoglio et al, these
proceedings) where the far-IR continuum is almost indistinguishable to
that of the archetypal starburst galaxy M82.

\section*{ACKNOWLEDGEMENTS}

DMA acknowledges PPARC for studentship support during this work and is
currently in receipt of an EC TMR network (FMRX-CT96-0068) postdoctoral
grant. AE acknowledges PPARC for postdoctoral support.


\begin{thebibliography}{}


\bibitem[\protect\astroncite{Alexander et~al.}{1999}]{aeh99}
Alexander, D.M., Efstathiou, A., Hough, J.H. et al.\ 1999, MNRAS, 310, 78
\bibitem[\protect\astroncite{Blanco et~al.}{1990}]{bla90}
Blanco, P.R., Ward, M.J., Wright, G.S., 1990, MNRAS, 242, 4P
\bibitem[\protect\astroncite{Bolton et~al.}{1949}]{bol49}
Bolton, J.G., Stanley, G.J., Slee, O.B., 1949, Nature, 164, 101
\bibitem[\protect\astroncite{Bruzual \& Charlot}{1993}]{bc93}
Bruzual, G., Charlot, S., 1993, ApJ, 405, 538
\bibitem[\protect\astroncite{Clarke et~al.}{1992}]{cla92}
Clarke, D.A., Burns, J.O., Norman, M.L., 1992, ApJ, 395, 444
\bibitem[\protect\astroncite{Eckart et~al.}{1990}]{eck90}
Eckart, A., Cameron, H., Rothermal, H. et al.\ 1990, ApJ, 363, 451
\bibitem[\protect\astroncite{Efstathiou et~al.}{1995}]{efs95}
Efstathiou, A., Hough, J.H., Young, S., 1995, MNRAS, 277, 1134
\bibitem[\protect\astroncite{Efstathiou et~al.}{2000}]{efs00}
Efstathiou, A., Rowan-Robinson, M., Siebenmorgan, R., 2000, MNRAS, 313, 734
\bibitem[\protect\astroncite{Hawarden et~al.}{1993}]{haw93}
Hawarden, T.G., Sandell, G., Matthews, H.E. et al.\ 1993, MNRAS, 260, 844
\bibitem[\protect\astroncite{Jaffe et~al.}{1993}]{jaf93}
Jaffe, W., Ford, H.C., Ferrarese, L. et al.\ 1993, Nature, 364, 213
\bibitem[\protect\astroncite{Jones et~al.}{1996}]{jon96}
Jones, D.L. et al.\ 1996, ApJ, 466, L63
\bibitem[\protect\astroncite{Marston \& Dickens}{1988}]{md88}
Marston, A.P., Dickens, R.J., 1988, A\&A, 193, 27
\bibitem[\protect\astroncite{Meadows \& Allen}{1992}]{ma92}
Meadows, V., Allen, D., 1992, Proc.ASA, 10(2), 104
\bibitem[\protect\astroncite{Morini et~al.}{1989}]{mor89}
Morini, M., Anselmo, F., Molteni, D., 1989, ApJ, 347, 750
\bibitem[\protect\astroncite{Mundel et~al.}{1995}]{mun95}
Mundel, C.G., Pedlar, A., Baum, S.A. et al.\ 1995, MNRAS, 272, 355
\bibitem[\protect\astroncite{Packham et~al.}{1993}]{pac96}
Packham, C., Hough, J.H., Young, S. et al.\ 1996, MNRAS, 278, 406
\bibitem[\protect\astroncite{Pier \& Krolik}{1992}]{pk92}
Pier, E., Krolik, J., 1992, ApJ, 401, 99
\bibitem[\protect\astroncite{Rowan-Robinson}{1992}]{r92}
Rowan-Robinson, M., 1992, MNRAS, 258, 787
\bibitem[\protect\astroncite{Rowan-Robinson \& Crawford}{1989}]{rc89}
Rowan-Robinson, M., Crawford, J., 1989, MNRAS, 238, 523
\bibitem[\protect\astroncite{Schreier et~al.}{1998}]{sch98}
Schreier, E.J., Marconi, A., Axon, D.J. et al.\ 1998, ApJ, 499, 143
\bibitem[\protect\astroncite{Tonry \& Schechter}{1990}]{ts90}
Tonry, J.L., Schechter, P.L., 1990, AJ, 100, 1794
\bibitem[\protect\astroncite{Wilson}{1988}]{w88}
Wilson, A.S., 1988, A\&A, 206, 41
\bibitem[\protect\astroncite{Young et~al.}{1996}]{yph96}
Young, S., Packham, C., Hough, J.H. et al.\ 1996, MNRAS, 283, L1

\end{thebibliography}
\end{document}